\documentclass[prd,preprint]{revtex4}

	\usepackage{amsmath}
	\usepackage{amsfonts}
    \usepackage{amssymb}
	\usepackage{makeidx}
	\usepackage[usenames,dvipsnames]{pstricks}
    \usepackage{subfigure}
	\usepackage{epsfig}
	\usepackage{pst-grad} 
	\usepackage{pst-plot} 
    \usepackage{graphicx}

\newcommand{\be}{\begin{equation}}
\newcommand{\ee}{\end{equation}}
\newcommand{\bea}{\begin{eqnarray}}
\newcommand{\eea}{\end{eqnarray}}

\makeindex

\begin{document}

\title{Low red-shift effects of local structure on the Hubble parameter in presence of a cosmological constant}
\author{Antonio Enea Romano${}^{1,2,3}$,Sergio Andres Vallejo${}^{3}$}

\affiliation{
${}^{1}$Department of Physics and CCTP, University of Crete, Heraklion 711 10, Greece \\
${}^{2}$Yukawa Institute for Theoretical Physics, Kyoto University, Kyoto 606-8502, Japan;\\
${}^{3}$Instituto de Fisica, Universidad de Antioquia, A.A.1226, Medellin, Colombia\\
} 

\begin{abstract}
In order to estimate the effects of local structure on the Hubble parameter we calculate the low-redshift expansion for $H(z)$ and $\frac{\delta H}{H}$  for an observer at the center of a spherically symmetric matter distribution in presence of a cosmological constant. We then test the accuracy of the formulae comparing them with fully relativistic non perturbative numerical calculations for different cases for the density profile.  The low red-shift expansion we obtain gives results more precise than perturbation theory since is based on the use of an exact solution of Einstein's field  equations. 
For larger density contrasts the low red-shift formulae accuracy improves respect to the perturbation theory accuracy because the latter is based on the assumption of a small density contrast, while the former does not rely on such assumption. 

The formulae can be used to take into account the effects on the Hubble expansion parameter due to the monopole component of the local structure.
If the $H(z)$ observations will show deviations from the $\Lambda CDM$ prediction compatible with the formulae we have derived, this could be considered an independent evidence of the existence of a local inhomogeneity, and the formulae could be used to determine the characteristics of this local structure.

\end{abstract}


\maketitle

\section{Introduction}
The standard cosmological model is based on the assumption that the Universe is homogeneous and isotropic on  sufficiently large scales. Nevertheless local observations could be strongly affected by local structure as shown for example in \cite{Romano:2014iea}, and it is important to study its  effects.
The analysis of luminosity density data \cite{Keenan:2013mfa} has provided some strong experimental evidence supporting the existence of  local inhomogeneities, but it would be important to confirm it using another observable such as the baryonic  acoustic oscillations (BAO) measurements \cite{Sahni:2014ooa,Li:2014yza,Blake:2012pj,Xu:2012fw,Samushia:2013yga,Font-Ribera:2013wce}.
The BAO scale allows in fact to determine the expansion rate of the Universe $H(z)$ independently from the luminosity distance and as such provides an important source of information about our Universe.
If the $H(z)$ estimations obtained from BAO observations data will show deviations from the $\Lambda CDM$ predictions  this could be considered an independent evidence of the existence of  local inhomogeneities.
This motivates the calculation of a low red-shift formula for $H(z)$, able to take into account  the effects of inhomogeneities which cannot be fully modeled with perturbation theory, as some of the inhomogeneities found for example in \cite{Keenan:2013mfa}.
A low red-shift expansion based on the use of exact solutions of Einstein's equations is in fact valid also for large values of the density contrast or of the gravitational potential.

The effects of a local inhomogeneity on cosmological observations have been studied already for different cases \cite{Romano:2014iea,Romano:2013kua,Clarkson:2010ej,Romano:2006yc,Romano:2009xw,Ben-Dayan:2014swa,Redlich:2014gga,Romano:2009qx,EneaRomano:2011aa,Marra:2011ct,Romano:2011mx,Romano:2010nc,Krasinski:2014zza,Romano:2012yq,Romano:2012ks,Balcerzak:2013kha,Romano:2012gk,2012GReGr..44..353R,Fanizza:2013doa,RomanoChen:2014,Krasinski:2014lna,Chung:2006xh,Romano:2014tya,Marra:2013rba} such as equation of state of dark energy or the luminosity distance \cite{Romano:2010nc,Romano:2012gk,Romano:2011mx}. 
It has been shown for example that the value of the cosmological constant could be affected significantly by the presence of  local inhomogeneity seeded by  
primordial curvature perturbations \cite{Romano:2013kua}, which could also lead to the wrong  conclusion of a varying equation of state for dark energy while only a cosmological constant is present \cite{Romano:2010nc}.
The origin of these effects is that spatial inhomogeneities can change the energy of the propagating photons, contaminating the cosmological red-shift due to the universe expansion, and consequently introducing some errors in the estimation of parameters based on cosmological models which ignore the effects of the inhomogeneities.

In this paper we will focus on the low red-shift effects of inhomogeneities on the Hubble expansion parameter, adopting an analytical approach based on the use of an exact solution of Einstein's equations to model the local structure.
We first derive the red-shift expansion of the geodesics equations, and  use it to obtain the expansion of $H(z)$. We then compute a formula for $\frac{\delta H}{H}$, the relative difference between the  $\Lambda CDM$ and the inhomogeneous case.  Finally we compare the formulae with the numerical calculations based on the integration of the Einstein's equation and the geodesics equations, finding a good agreement. We also check that the low red-shift expansion formulae are more precise than the perturbative calculation, especially when the density contrast is larger.

\section{Modeling the local Universe}
We use the LTB solution  to model the monopole component of the  local structure \cite{Lemaitre:1933,Lemaitre:1933qeLemai,Lemaitre:1931zz,Tolman:1934za,Bondi:1947av} 
\begin{eqnarray}
\label{LTBmetric} %
ds^2 = -dt^2  + \frac{\left(R'(t,r)\right)^2 dr^2}{1 + 2\,E(r)}+R(t,r)^2
d\Omega^2 \,, 
\end{eqnarray}
where $R$ is a function of the time coordinate $t$ and the radial
coordinate $r$, $E(r)$ is an arbitrary function of $r$, and
$R'(t,r)=\partial_rR(t,r)$. 
The Einstein's equations imply that 
\begin{eqnarray}
\label{eq2} \left({\frac{\dot{R}}{R}}\right)^2&=&\frac{2
E(r)}{R^2}+\frac{2M(r)}{R^3}+\frac{\Lambda}{3} \,,  \\
\label{eq3} \rho(t,r)&=&\frac{2M'}{R^2 R'} \,, 
\end{eqnarray}
where $M(r)$ is an arbitrary function of $r$, $\dot
R=\partial_tR(t,r)$ and we choose a system of units in which $c=8\pi G=1$. 

To compute $H(z)$ we need to solve the radial null geodesics \cite{Celerier:1999hp} 
\begin{eqnarray}
{dr\over dz}&=&{\sqrt{1+2E(r(z))}\over {(1+z)\dot {R'}[t(z),r(z)]}} \,,
\label{eq:34} \\
{dt\over dz}&=&-\,{R'[t(z),r(z)]\over {(1+z)\dot {R'}[t(z),r(z)]}} \,, 
\label{eq:35} 
\end{eqnarray}
and then we substitute in the formula for the Hubble parameter in a LTB space \cite{Chatterjee:2011,Bellido:2009}
\bea
H(t,r)&=&\frac{2}{3} H_{\perp}(t,r) + \frac{1}{3} H_{\|}(t,r) \,, \\
H(z)&=&H(t(z),r(z))\,,
\eea
where
\bea
H_{\perp}(t,r)\equiv \frac{\dot{R}(t,r)}{R(t,r)}  \,, \\
H_{\|}(t,r)\equiv\frac{\dot{R}' (t,r)}{R' (t,r)} \,.
\eea
The analytical solution can be derived \cite{Zecca:2013wda,Edwards01081972} introducing a new coordinate $\eta=\eta(t,r)$,  and new functions $\rho_0(r)$ and $k(r)$ given by
\bea
\frac{\partial \eta}{\partial t}|_r&=& \frac{r}{R}=\frac{1}{a}\,, \\
\rho_0(r)&=&\frac{6 M(r)}{r^3}\,, \\
k(r)&=&-\frac{2E(r)}{r^2}\,.
\eea
We will adopt, without loss of generality, the coordinate system in which $\rho_0(r)$ is a constant, the so called FLRW gauge. 
We can then express Eq. (2) in the form
\begin{equation}
\left(\frac{\partial a}{\partial \eta}\right)^2= -k(r)a^2 + \frac{\rho_0}{3}a + \frac{\Lambda}{3}a^4 \,,
\end{equation}
The coordinate $\eta$, which can be considered a generalization of the conformal time in a homogeneous FLRW universe, is defined implicitly by Eq. (10). The relation between $t$ and $\eta$ is obtained by integrating of Eq. (10) and is  given by \cite{RomanoChen:2014}
\begin{equation}
t(\eta,r)=\displaystyle\int_{0}^{\eta} a(x,r) \, dx +t_b(r) \,,
\end{equation}
where $t_b(r)$ is a functional constant of integration called bang function, since it corresponds to the fact that in these models the scalar factor can vanish at different times at different locations. We will consider models with $t_b(r)=0$.
The solution of eq.(13) can then be written in the form
\begin{equation}
a(\eta,r)=\frac{\rho_0}{k(r)+3 \wp (\frac{\eta}{2};g_2(r),g_3(r))} \,, \label{aeta} 
\end{equation}
where $\wp(x;g_2,g_3)$ is the Weierstrass elliptic function and
\bea
g_2(r)=\frac{4}{3}k(r)^2 \,,\quad
g_3(r)=\frac{4}{27} \left(2k(r)^3 -\Lambda\rho_0^2\right)\,.
\eea
In terms of $\eta$ and $a(\eta,r)$ the radial null geodesics and the Hubble parameter are given by \cite{Romano:2009xw}
\bea
\frac{d\eta}{dz}&=&-\frac{\partial_r t(\eta,r) + G(\eta,r)}{(1+z)\partial_\eta G(\eta,r)} \,, \\
\frac{dr}{dz}&=&\frac{a(\eta,r)}{(1+z)\partial_\eta G(\eta,r)} \,,\\
H(\eta,r)&=& H(t(\eta,r),r) \,, \label{Heta}
\eea
where
\begin{equation}
G(\eta,r)\equiv\frac{R,_r}{\sqrt{1+2E(r)}} =\frac{[\partial_r (a(\eta,r)r) - a^{-1} \partial_\eta (a(\eta,r)r) \partial_r t(\eta,r)]}{\sqrt{1-k(r)r^2}} \,.
\end{equation}
The function $G(\eta,r)$ has an explicit analytical form, making the above geodesics equations particularly suitable for a low-red-shift expansion. 

\section{Low red-shift expansion of the  Hubble parameter $H(z)$}
In order to obtain a low-redshift formula for the Hubble parameter we   expand  the function $k(r)$ as
\begin{align}
k(r)&=k_0 + k_1 r + k_2 r^{2} + ... \,,
\end{align}
We also need an expansion for $t(\eta,r)$ which can be obtained from the exact solution for $a(\eta,r)$ according to 
\begin{equation}
t(\eta,r)=t_0(r)+a(\eta_0,r)(\eta-\eta_0)+\frac{1}{2}\partial_{\eta}a(\eta_0,r)(\eta-\eta_o)^2+... \,,
\end{equation}
where we defined the function $t_0(r)$ by
\begin{equation}
t_0(r)\equiv t(\eta_0,r) \,.
\end{equation}
Using the properties of the Weierstrass elliptic functions $\wp$, $\zeta$ and $\sigma$ \cite{Abramowitz:1972aa}  we can compute the integral in eq.(14), obtaining 
\begin{equation}
t(\eta,r)=\frac{2 \rho _0}{3 \wp '\left(\wp
   ^{-1} \left( {-\frac{k(r)}{3}} \right)\right)} \left[ \ln \left(\frac{\sigma \left(\frac{\eta
   }{2}-\wp ^{-1}\left(-\frac{k(r)}{3}\right)\right)}{\sigma \left(\frac{\eta}{2}+\wp ^{-1}\left(-\frac{k(r)}{3}\right)\right)}\right)+\eta  \zeta \left(\wp
   ^{-1}\left(-\frac{k(r)}{3}\right)\right) \right] \,,
\end{equation}
where $\wp '$ is the derivative of the Weierstrass' elliptic function $\wp$. $\wp^{-1}$, $\zeta$ and $\sigma$ are defined by the equations: 
\begin{align}
\wp^{-1}\left(\wp\left(x\right)\right) &= x \,, \\
\zeta ' \left(x\right) &= - \wp \left(x \right) \,, \\
\frac{\sigma ' \left(x \right)}{\sigma \left(x \right)} &= \zeta \left(x \right) \,. 
\end{align}
The use of the exact expression for $t(\eta,r)$ improves the accuracy for the expansion for the geodesics respect to previous calculations \cite{Romano:2012gk}, which were based  on a pertubative expansion of $t_0(r)$, rather than the use of the exact value.

Now we can find the low red-shift Taylor expansion for the geodesic equations \cite{Romano:2010nc}, and then calculate the Hubble parameter.
We  expand the solution of the geodesic equations according to
\bea
r(z)&=&r_1z+r_2z^2+r_3z^3 + ...\\
\eta(z)&=&\eta_0 +\eta_1 z +\eta_2 z^2+...
\eea
After substituting the above expansion in the geodesic equations we can map the solution of the system of differential equations into the solution of a system of algebraic equations for the coefficients of the expansion. 
Here we give the formulae for the case in which $k_{0}=0$, which is enough to understand qualitatively the effects of the inhomogeneity.
The term $k_{0}$ corresponds in fact to the homogeneous component of the curvature function, which in absence of inhomogeneities is simply the curvature of a FLRW model, and as such is not associated to any new physical effect not already known from standard cosmology. 

For the geodesics we get:  
\begin{align}
\eta _1 &= -\frac{1}{a_0^2 H_0} [a_0+ t_0'(0)] \,, \\
\eta _2 &= \frac{1}{12 a_0^3 H_0^2 \Omega _M} [ a_0 H_0 t_0'(0) \left(3 \Omega _M \left(9 \Omega _M-4\right)-8 K_1\right)+a_0^2 H_0 \left(9 \Omega _M^2-4 K_1\right)+ \nonumber \\
&\quad {} +6 \Omega _M \left(3 H_0
   \left(\Omega _M-1\right) t_0'(0)^2-t_0''(0)\right) ] \,, \\
r_1 &= \frac{1}{a_0 H_0} \,, \\
r_2 &= \frac{1}{12 a_0^2 H_0 \Omega _M} [ a_0 \left(4 K_1-9 \Omega _M^2\right)+6 \left(2-3 \Omega _M\right) \Omega _M t_0'(0) ] \,, \\ 
r_3 &= \frac{1}{72 a_0^3 H_0^2 \Omega _{\Lambda } \Omega _M^2} [ a_0^2 H_0 \left(4 K_1^2 \left(2 \Omega _{\Lambda }+\zeta _0 \left(2-3 \Omega _M\right)+\Omega _M\right)-60 K_1 \Omega _{\Lambda } \Omega _M^2 + \nonumber \right. \\ & \quad {} \left. +3
   \Omega _{\Lambda } \Omega _M \left(8 K_2 +3 \left(9 \Omega _M-4\right) \Omega _M^2\right)\right)-36 a_0 H_0 \Omega _{\Lambda } \Omega _M \text{t0}'(0)
   \left(K_1 \left(4 \Omega _M-2\right)+ \nonumber \right. \\ & \quad {} \left. +\left(8-9 \Omega _M\right) \Omega _M^2\right)+ 18 \Omega _{\Lambda } \Omega _M^2 \left(6 H_0 \left(3 \Omega _M^2-4
   \Omega _M+1\right) t_0'(0)^2+\left(2-3 \Omega _M\right) t_0''(0)\right) ] \,,
\end{align}
where $\Omega_M$, $\Omega_{\Lambda}$, $T_0$ and $K_n$ are dimensionless quantities given by  \cite{EneaRomano:2011aa} 
\bea
\rho_0&=&3 \Omega_M a_0^3 H_0^2 \,, \\
\Lambda&=&  3 \Omega_{\Lambda} H_0^2 \,, \\
T_0&=& \eta_0 \left(a_0 H_0 \right) \,, \\
K_n&=& k_n (a_0 H_0)^{n+2} \,.
\eea
We have also used the following definitions 
\bea
a_0 & = &a (\eta_0,0) \,, \\
H_0 & = & H(\eta_0,0) \,, \\
\zeta_0 & = & \zeta \left(\frac{T_0}{2};\frac{4 K_0^2}{3},\frac{4}{27}\left(2 K_0^3 - 27 \Omega_{\Lambda} \Omega_M^2\right)\right) \,,
\eea
where $\zeta$ is the Weierstrass Zeta Function.
As we can see in the above formulae the effects of the inhomogeneity start to show respectively at first order for $\eta(z)$ and second order for $r(z)$.

\begin{figure}[ht]
\centering
\includegraphics[width=0.8\textwidth]{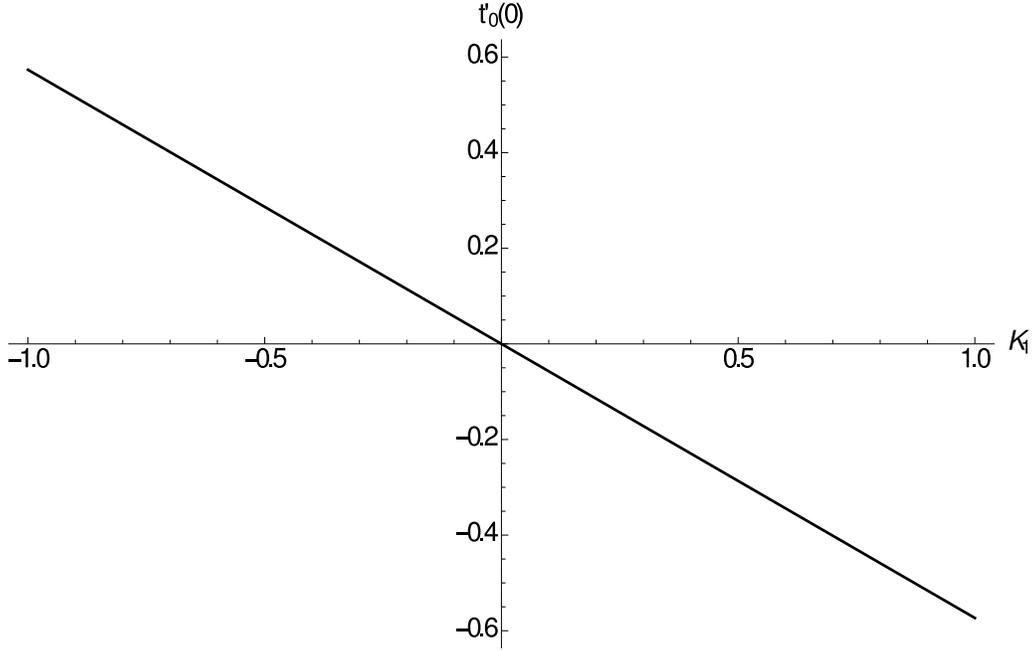}
\caption{We plot $t_0'(0)$ as a function of $K_1$. This is the quantity determining  the leading order effect for $\frac{\delta H}{H}$ as shown in eq.(\ref{dh1}).}
\label{tp0}
\end{figure}

\begin{figure}[ht]
\centering
\includegraphics[width=0.8\textwidth]{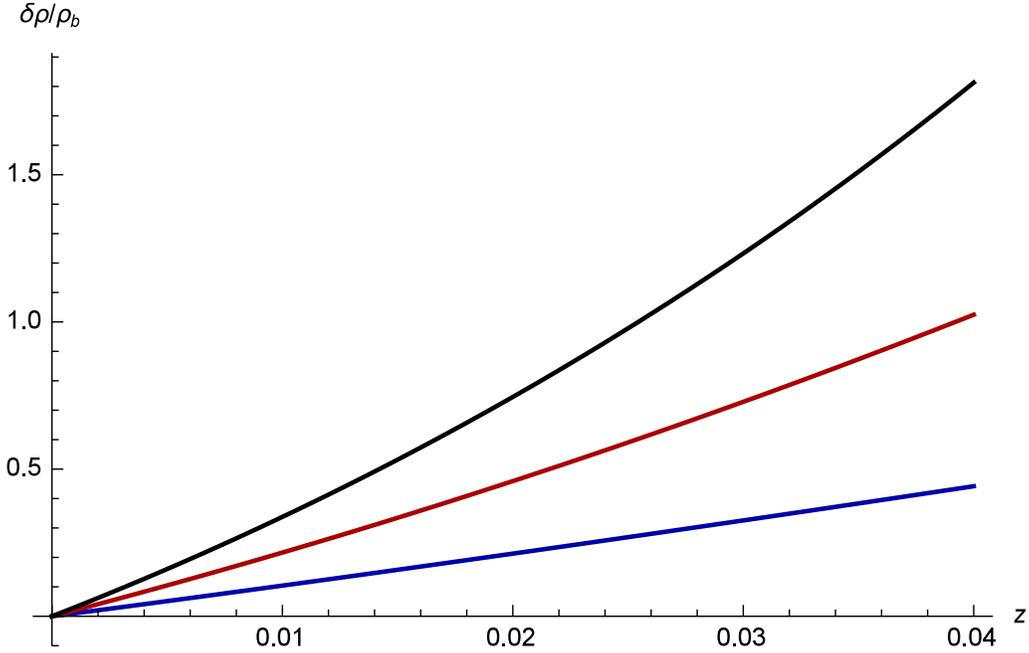} \ 
 \caption{The density contrast $\delta =\frac{\delta \rho }{{\rho_b}}$ is plotted as a function of the redshift for three different inhomogeneities profiles modeled by LTB solutions.}
\label{drho}
\end{figure}

In order to obtain a formula for the Hubble parameter as a function of the red-shift we need to substitute eq.(17-18) in eq. (\ref{Heta})
\bea
H(z)&=&H(\eta(z),r(z)) \,. \label{Hltb}
\eea
After expanding up to second order in $z$ we get:
\begin{align}
H(z)&=H_0+H_1 z + H_2 z^2 \,, \\
H_1&=\frac{1}{2} H_0 \Omega _M \left(\frac{4 t_0'(0)}{a_0}+3\right) \,, \\ 
H_2&=\frac{1}{72 a_0^2 \Omega _{\Lambda } \Omega _M} [a_0^2 H_0 \left(20 \left(\zeta _0-1\right) K_1^2+48 K_1 \Omega _{\Lambda } \Omega _M+27 \Omega _{\Lambda } \left(4-3 \Omega _M\right) \Omega
   _M^2\right)+ \nonumber \\ & \quad {} +6 a_0 H_0 \Omega _{\Lambda } \Omega _M t_0'(0) \left(20 K_1+9 \left(8-5 \Omega _M\right) \Omega _M\right) +18 \Omega _{\Lambda }
   \Omega _M^2 \left(H_0 \left(25+ \right. \right. \nonumber \\ & \quad \left. \left. -12 \Omega _M\right) t_0'(0)^2+5 t_0''(0)\right) ] \,.
\end{align}
The procedure to reduce the analytical formula to this form is rather complicated since it involves to express wherever possible all the intermediate expressions in terms of physically meaningful quantities using the properties of the Weierestrass elliptic functions \cite{Abramowitz:1972aa} and  we give more details in  appendix A.

As we can see from the first order coefficient $H_1$,  at leading order in   $t_0'(0)$ determines the sign of the correction respect to the homogeneous case, and for this reason we plot $t_0'(0)$ as a function of $K_1$ in fig.(\ref{tp0}).
At second order we have a more complicated dependency for $H_2$, which involves also $K_2$ and $t''_0(r)$.

We can easily interpret the linear behavior shown fig. 1, applying the chain rule for the derivative 
\bea
t_0'(0)&=&\frac{\partial t_0(r) }{\partial k}\frac{\partial k}{\partial r}|_{r=0}=\alpha K_1 \,,\\
\alpha&=& (a_0 H_0)^{-3} \frac{\partial t_0(r) }{\partial k}|_{k=k_0}\approx-0.57 \,. \label{alpha}
\eea




\section{Relative difference of $H(z)$ respect to the homogeneous case}
For a flat FLRW solution the expansion rate is given by
\begin{equation}
H^{FLRW}(z)= H_0 \sqrt{\Omega _M (1+z)^3 + \Omega _{\Lambda}} \,.
\end{equation}
Since we want to compare the inhomogeneous case with the flat  FLRW case we define the relative difference as
\begin{equation}
\frac{\delta H (z)}{H} =  \frac{H^{\Lambda LTB}(z)}{ H^{FLRW}(z)} - 1 \,,
\end{equation}
where we are denoting with $H^{\Lambda LTB}(z)$  the expansion rate defined in eq.(\ref{Hltb}).

We can now expand the above expression at low-redshift to get
\begin{align}
\frac{\delta H (z)}{H} &= \frac{\delta H_{1}}{H} z + \frac{\delta H_{2}}{H} z^2 + ... \,, \\ 
\frac{\delta H_{1}}{H} &=\frac{2 \Omega _M t_{0}'(0)}{a_0}=\frac{2\alpha\Omega_M}{a_0}K_1 \,, \label{dh1} \\
\frac{\delta H_{2}}{H} &=\frac{1}{36 a_0^2 H_0 \Omega _{\Lambda } \Omega _M} [  2 a_0^2 H_0 K_1 \left(5 \left(\zeta _0-1\right) K_1+12 \Omega _{\Lambda } \Omega _M\right)+3 a_0 H_0 \Omega _{\Lambda } \Omega _M
   t_{0}'(0) \left(20 K_1 + \right. \nonumber \\ & \quad {} \left.  +9 \left(8-9 \Omega _M\right) \Omega _M\right)+9 \Omega _{\Lambda } \Omega _M^2 \left(H_0 \left(25-12 \Omega _M\right)
   t_{0}'(0)^2+5 t_{0}''(0)\right) ] \,. \label{dhz}
\end{align}
It is  straightforward to verify that in the homogeneous limit, when $K_0=K_1=K_2=0$ , as expected, $\delta H_1 = \delta H_2  =0$, since  $t_0'(0) = t_0''(0) = 0$ .
From the expression for $\frac{\delta H_{1}}{H}$ it is clear the crucial role played by $t_0'(0)$, which determines the sign of the relative difference at leading order in red shift, and according to eq.(46-47) is proportional to $K_1$.

\begin{figure}[ht]
\centering
\includegraphics[width=.9\textwidth]{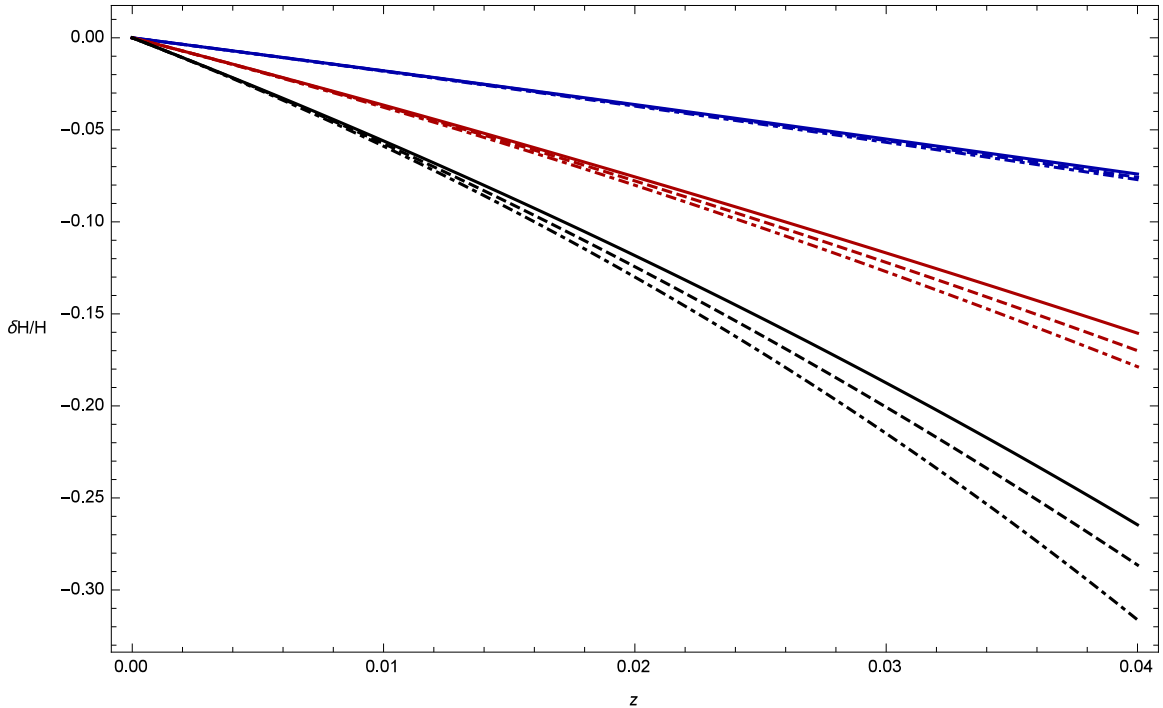} \ 
\caption{The relative difference with respect to the homogeneous case $\frac{\delta H(z)}{H}$ is plotted as a function of the redshift for three different density contrasts. The colors correspond to the density contrasts in fig.(\ref{drho}). The solid lines are for the numerical  calculation, the dashed lines are for the analytical formulae and the dot-dashed lines are for the perturbation theory result.}
\label{dhm}
\end{figure}

\begin{figure}[ht]
\centering
\includegraphics[width=.9\textwidth]{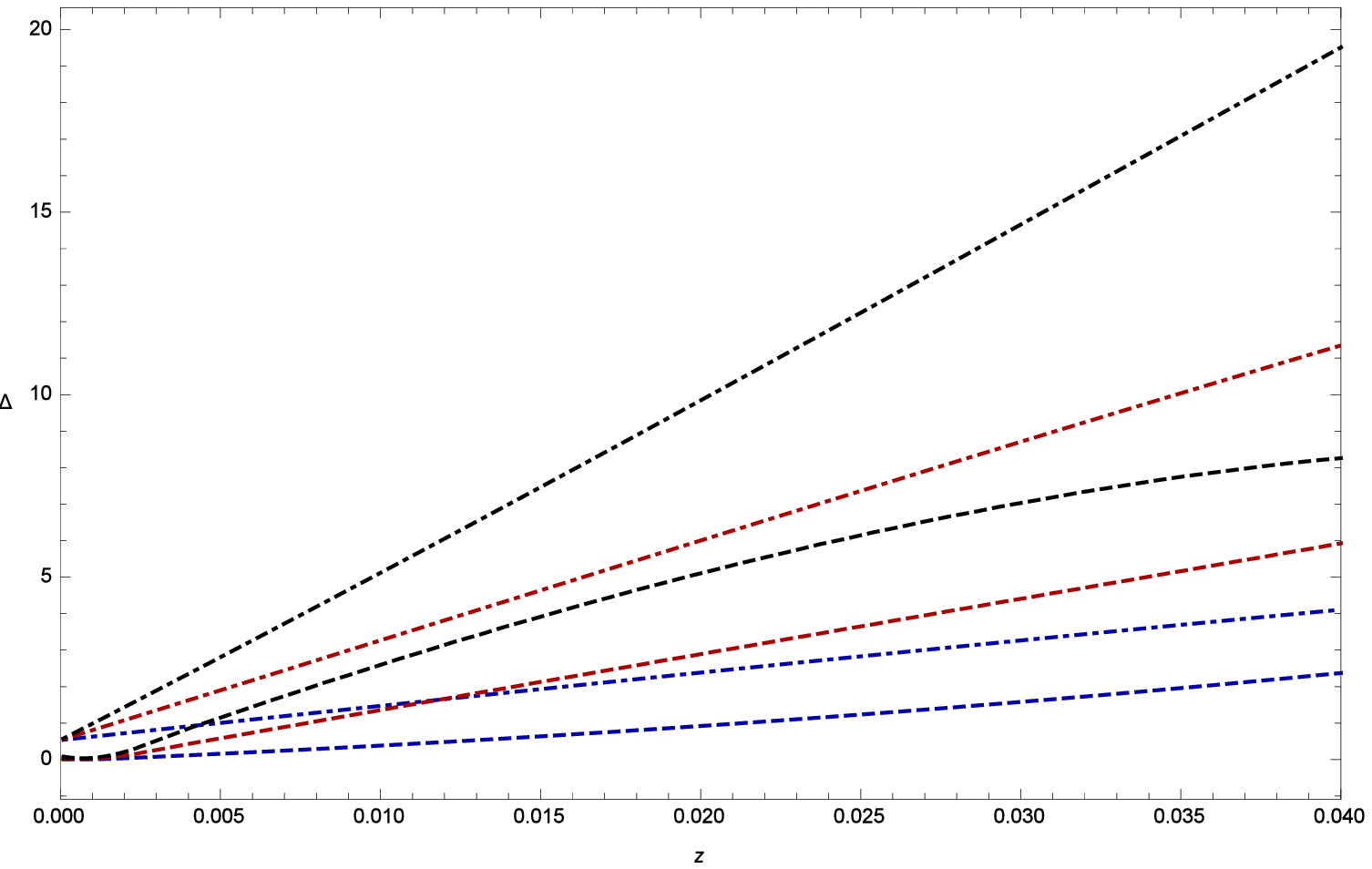}
\caption{The relative percentual difference $\Delta(z) = 100 \frac{\delta H/H-(\delta H/H) ^{num}}{(\delta H /H)^{num}}$ of $\delta H/H$  respect to the numerical results is plotted as a function of the redshift for the analytical formula (dashed) and for perturbation theory (dot-dashed). The colors correspond to the density contrasts in fig.(\ref{drho}). As it can be seen the low red-shift formula is always better, especially for larger density contrasts.} 
\label{dhp}
\end{figure}

Using eq.(\ref{alpha}) we can also see that for a fixed $K_0$ the sign of $K_1$ determines the sign of $\frac{\delta H_{1}}{H}$ at very low red-shift when the second order contributions can be neglected.
This in in good agreement with fig.(\ref{dhm}-\ref{dhp}), which show an approximate linear behavior with opposite slope, corresponding to different values of $K_1$.

\subsection{Testing the accuracy of the formulae}
In order to test the accuracy of the analytical formulae and compare it with perturbation theory we  perform numerical calculations using LTB solutions corresponding to the density contrasts shown in fig.(\ref{drho}).
Using large density contrasts we can test the limit of the perturbation theory results and compare them with the low red-shift expansion. 

As seen in fig.(\ref{dhm}-\ref{dhp})  the analytical formula for the relative difference of $H(z)$ respect to the homogeneous case is in good agreement with $\frac{\delta H(z)}{H}$ obtained by integrating numerically the geodesics and background equations and is more accurate than the perturbation theory.
From fig.(\ref{dhp}) in particular we can see that the agreement with the exact numerical calculation for the red-shift expansion is in general better than that of the perturbation theory result. For larger density contrasts, as expected, the pertubative calculation is increasingly inaccurate while the red-shit expansion is still in good agreement with the exact numerical result, since it does not rely on the assumption of a small density contrast. 
This is due to the fact that perturbation theory is based on the assumption that $\delta\rho/\rho \ll 1$, while the low red-shift expansion is based on an exact solution of the Einstein's equations.

\section{Conclusions}
We have derived a low-redshift analytical formula for the Hubble parameter for an observer at the center of a  spherically symmetric matter distribution, using an exact solution of Einstein's field equations.
Such a formula is in good agreement with numerical calculations and is more accurate than the perturbation theory result, especially for large density contrasts.
This is due to the fact that perturbation theory is based on the assumption that $\delta\rho/\rho \ll 1$, while the low red-shift expansion is based on an exact solution of the Einstein's equations. 

If the $H(z)$ observations will show deviations from the $\Lambda CDM$ predictions compatible with the formulae we have derived, this could be considered an independent evidence of the existence of a local inhomogeneity, and the formulae could be used to determine the characteristics of this local structure.

While the expansion for $r(z)$ depends on our choice of radial coordinate, the formulae for the $H(z)$ are independent of it, since both $H$ and the redshift are physically observable quantities, and as such are independent of the gauge choice.
Since in the  derivation of the formulae the
inhomogeneity profile is determined by the coefficients of the expansion of the function $K(r)$, there is still some coordinate dependency left in the way we parameterize the inhomogeneity.
While our choice of gauge is quite natural, since in the $FRW$ gauge, corresponding to $\rho_0(r)=const$, the radial coordinates reduces to the radial $FRW$ comoving coordinate in the limit in which the function $k(r)$ goes to zero, a totally coordinate independent formula would still be preferable.
For this reason in the future it will  be interesting to derive formulae which are completely  independent of the choice of the coordinate system, parameterizing the inhomogeneity in terms of the redshift expansion of the density $\rho(z)$, which is a physical observable, instead of using the expansion of $K(r)$.

\appendix
\section{ Derivation of the analytical formula}
In order to obtain the formula for the red-shift expansion of $H(z)$ we have applied several manipulations and substitutions. 
The method is based on re-expressing everything in terms of physical quantities, starting from the definitions of $a_0$ and $H_0$, which are related to $\wp$ and $\wp '$ by the equations
\begin{align}
a_0 & \equiv  (\eta_0,0) = \frac{\rho _0}{k_0 + 3 \wp _0} \,, \\
H_0 & \equiv  H(\eta_0,0) = - \frac{3 \wp' _0}{2 \rho _0} \,, \\
\end{align}
where 
\begin{align}
\wp _0 &= \wp (\eta _0; g_2(0),g_3(0)) \,, \\
\wp' _0 &= \frac{\partial \wp (\eta; g_2(0),g_3(0)) }{\partial \eta} |_{\eta=\eta _0} \,.
\end{align}
By inverting the previous equations we obtain the following relations
\begin{align}
\wp _0 &= \wp (\eta _0; g_2(0),g_3(0)) = \frac{\rho _0 - a_0 k_0}{3 a _0} \,,\\
\wp' _0 &= \frac{\partial \wp (\eta; g_2(0),g_3(0)) }{\partial \eta} |_{\eta=\eta _0} = - \frac{2 H_0 \rho _0}{3} \,.
\end{align}
We can then substitute the above expressions everywhere $\wp$ and $\wp '$ appear, making  the final formula only depending on physical quantities such as $H_0$.
In order to simplify the results we have also used the Einstein's equation at the center $(\eta _0 ,0)$ 
\begin{equation}
1=-K_0+\Omega _M + \Omega _{\Lambda} \,.
\end{equation}

\acknowledgments
AER thanks  Ryan Keenan and Licia Verde for useful comments and discussions, the Department of Physics of Heidelberg University for the kind  hospitality, and Luca Amendola, Valeria Pettorino and Miguel Zumalacarregui for insightful discussions. 

This work was supported by the European Union (European Social Fund, ESF) and Greek national funds under the ``ARISTEIA II'' Action, the Dedicacion exclusica and Sostenibilidad programs
at UDEA, the UDEA CODI
projects IN10219CE and 2015-4044.



\bibliographystyle{h-physrev4}
\bibliography{mybib}
\end{document}